\documentclass[sigconf]{acmart} %  anonymous, review
\usepackage{booktabs} % For formal tables
\usepackage{standalone}
\usepackage{tabu}
\setcopyright{rightsretained}
\acmDOI{10.475/123_4}

\acmISBN{123-4567-24-567/08/06}

\copyrightyear{2017} 
\acmYear{2017} 
\setcopyright{acmcopyright}
\acmConference{CIKM'17 }{November 6--10, 2017}{Singapore, Singapore}\acmPrice{15.00}\acmDOI{10.1145/3132847.3133101}
\acmISBN{978-1-4503-4918-5/17/11}

\begin{document}
\title{Smart City Analytics: Ensemble-Learned Prediction of Citizen Home Care}

\author{Casper Hansen, Christian Hansen, Stephen Alstrup, and Christina Lioma}
\affiliation{%
  \institution{Department of Computer Science, University of Copenhagen, Denmark}
}
\email{{c.hansen, chrh, s.alstrup, c.lioma}@di.ku.dk}

%\renewcommand{\shortauthors}{C. Hansen et al.}

\begin{abstract}
We present an ensemble learning method that predicts large increases in the hours of home care received by citizens. The method is supervised, and uses different ensembles of either linear (logistic regression) or non-linear (random forests) classifiers. Experiments with data available from 2013 to 2017 for every citizen in Copenhagen receiving home care (27,775 citizens) show that prediction can achieve state of the art performance as reported in similar health related domains (AUC=0.715). We further find that competitive results can be obtained by using  limited information for training, which is very useful when full records are not accessible or available. Smart city analytics does not necessarily require full city records. 

To our knowledge this preliminary study is the first to predict large increases in home care for smart city analytics.
\end{abstract}

%\begin{CCSXML} 
%\end{CCSXML}
%\ccsdesc[500]{is this needed~is this needed?}
\maketitle
\section{Introduction and Related Work}
In many countries, older citizens are a rapidly growing proportion of the population, and many of them need varying levels of home care, such as help with medication or surgery rehabilitation, for instance. %Different factors influence the individual home care level needed, but 
Predicting how much home care will be needed, and by whom, is important for at least two reasons: (1) it allows for preventive measures to be taken in order to forestall or curb health deterioration, leading to better quality of life for citizens; (2) it facilitates better planning of the city's resources, leading to reducing the overall financial cost of healthcare to the city, while also improving efficiency in response time and waiting lists. Both of the above are necessary actions towards creating \textit{smart cities}. %who will need large amounts of additional care ahead of time can allow preventive measures to be initiated faster, thus attempting to avoid the financial cost associated with the increased amount of home care, and positively influencing the citizen. Similarly, it can be useful for planning purposes. In Copenhagen, the capital of Denmark, the assessment of the home care need is done manually, but with more than 27,000 citizens receiving home care since 2013, predictions on who will need much more home care would be beneficial, and further the progress towards being a \textit{Smart City}.

We present a method for automatically predicting citizen home care. Specifically, our method uses ensemble learning to predict large increases in the amount of home care needed by citizens. 
The method creates a new prediction model each month using 3 months of prior information, and the collection of models is combined to a single prediction through meta learning. We focus on predicting large increases as those developments are the most costly and hard to manually predict. 

%We have detailed daily logs available on all citizens in Copenhagen receiving home care, which we study on five cumulative information levels to determine what performance can be expected depending on the used logging.

To the best of our knowledge no research has been done on predicting large long term increases in home care on the individual level, and generally research on home care prediction is scarce. Lanzarone et al. \cite{lanzarone2010patient} proposed a stochastic model to represent a citizens' care pathway, and based on historical data they trained and evaluated the models ability on predicting the number of visits 1 week ahead. Their focus was on providing a support system in the short term, while our work focuses on predicting long term increases before they happen. Generally, a large increase in home care corresponds to a decrease in a citizens' capabilities, either through natural causes such as ageing or from sudden illness. As such, this line of research is similar to that of predicting hospital re-admission, where recent model comparison studies achieve AUC scores ranging from 0.68 to 0.73 \cite{tong2016comparison, futoma2015comparison} depending on the used data.

%The central research questions addressed in this paper is 1) how well can large increases in home care be predicted on the individual citizens, and 2) what level of detail in the data is needed for that purpose.

Our work contributes a new application of ensemble learning for smart city analytics, which can predict large increases in citizen home care with high accuracy and robustly across reduced amounts and types of training data.
\section{Ensemble Learning for Home Care Prediction}
\label{methodd}
%\input{../method/method.tex}
%** REWRITE THIS SECTION FOLLOWING THE FORMAT OF SECTION 2 IN \url{https://arxiv.org/pdf/1609.00689.pdf}. FOR EXAMPLE, WE NEED FORMULAS TO DESCRIBE THE LEVELS OF MODELS BEING COMBINED, AND A CLEAR DESCRIPTION OF THE COMBINATION AND ANY PARAMETERS THAT ARE INVOLVED.**
%
%** FIRST WE SHOULD PRESENT THE MODEL. THEN WE SHOULD SAY HOW WE USE THIS MODEL TO STUDY IF NON-STATIONARITY AFFECTS THE PREDICTION. THESE CAN BE TWO SEPARATE SUBSECTIONS.**

We formulate predicting large increases in home care as a binary classification problem. Given a citizen's past, the task is to predict whether the number of provided home care hours goes up by at least 6 hours in the next 3 months. We do the prediction using ensemble learning, which consists of forming an ensemble of level 0 models trained over time, and then combining the level 0 predictions into a single prediction using a level 1 meta model. We consider data to be separated in chunks, such that at time $t$ a training dataset $D^t$ arrives as instances (vectors) $x^t(i)$ and target variables $y^t(i) \in \{-1,1\}$ for $i=1,...,|D^t|$, where $x^t(i)$ and $y^t(i)$ are sampled from an unknown distribution, which is potentially changing at each time step. At each time step $t$ we train a model $h_t$, such that we at a given $t$ have an ensemble of level 0 models $\{h_1,...,h_t\}$. To combine the prediction of each level 0 model, we build a new set of instances by computing $x^t_h(i) = [h_1(x^t(i)),...,h_t(x^t(i))]^T$ for all $i$, i.e. a vector consisting of each level 0 model's prediction for each instance. These instances are used to built the final level 1 model. 

We experiment with two types of models for both level 0 and level 1. The first is a linear logistic regression model, and the second is a nonlinear Random Forest model \cite{breiman2001random}. 
Training either level 0 or level 1 models is done notationally in the same way when we use the notation defined above, so when describing the models we simply refer to a dataset consisting of $(x,y)$ pairs.

\textbf{Logistic regression} in its regularized form is defined as the minimization of:
\begin{equation}
\min_{\vec{w}} \sum_{x,y} \log(1+\exp(-\vec{w}^Tx \cdot y)) + \lambda ||\vec{w}||^2
\end{equation}
consisting of a weight parameter $\vec{w}$ and a regularizations parameter $\lambda$. $\vec{w}_0$ denotes a bias term such that %we prepend 1 to $x$ such that 
$x_0=1$. 

\textbf{Random Forest} is an ensemble method and consists of a combination of weak classifiers in the form of tree predictors. It consists of three parameters \cite{breiman2001random}: the number of trees, the number of features to consider in each split, and the minimum number of samples for splitting and creating leaves. %We explain their tuning in Section \ref{s:exp}.

The rationale of experimenting with these two models is that, while logistic regression is only able to capture linear relationships, it is a less complex model and thus requires fewer samples to train in order to expect good performance. This may be beneficial since data is not plentiful in this domain. Random Forest is able to capture nonlinear relationships and was recently empirically shown in general to be among the best families of classifiers \cite{rfbest}. However, it is a more complex model thus requiring more samples to train than logistic regression. 

Ensemble learning methods \cite{de2,krawczyk2017ensemble} focus on how to combine the level 0 models to obtain the final prediction, and most often do not assume a specific level 0 model. Different combination strategies exist \cite{krawczyk2017ensemble}, and we have employed the \textit{dynamic} strategy where the learned level 0 models are kept unchanged over time, and the ensemble adapts via a combination phase where a weighting or selection of the level 0 models is learned. Our approach resembles the Dynamic Ensemble of Ensembles (DE$^2$) method \cite{de2}, that also employs interim ensembles of the level 0 models.%, but DE$^2$ uses the same base learner in each level. However, we experiment with different base learners for each step, as the complexity of the task in each level is very different, and thus using the same base learner may not be optimal.

\section{Experimental Evaluation}
\label{s:exp}

\subsection{Data}
We evaluate our approach on the following data provided by the city of Copenhagen: from April 2013 to April 2017 (inclusive), records of all citizens of Copenhagen who have received home care (27,775 citizens in total). We adopt the definition of home care used by the city, which covers a wide variety of services, spanning from daily help with e.g. pills and laundry, to rehabilitation at home after surgery. The contents of this data are displayed in Table \ref{tab:data}. The feedback is reported by the care taker after each visit, and potential hospitalization is reported by the hospitals. The provided services are tracked on an individual basis.

%, in the rest of the paper each service they receive will be termed a home care service. On each home care service provided we have extensive information describing it with regards to when it was provided, how long it took, what kind of service it was, and its price. We also have access to simple care taker feedback with regards to if the citizen was home or hospitalized. In addition we have access to information on each citizen with regards to gender, age, civil status, and housing information. These information sources are summarized in categories in Table \ref{tab:data}.

\begin{table}[!ht]
\vspace{-0pt}
\caption{Our dataset of home care citizen records in Copenhagen from 2013 - 2017.}
\vspace{-10pt}
\label{tab:data}
\begin{minipage}{\linewidth}
\begin{center}
\begin{tabu}{lX}
\toprule
\textbf{Category} & \textbf{Features} \\
Basic & gender, age, zipcode, date, civil status of citizen\\
Living type & own residence, senior housing, assigned residence \\
Time & day, evening, night, weekday, weekend \\
Type & public, private \\
Health Care & generic activity, emergency care stay, dementia care, reoccurring visit, dental care, palliative care, personal care, practical help, rehabilitation, sick care \\
Feedback & citizen home, citizen not home, citizen hospitalized, other \\
Length & number of home care hours, number of large increases of at least 6 hours within the last 3 months \\
Financial & financial cost of service\\
\bottomrule
\end{tabu}
\end{center}
\end{minipage}
\end{table}
\begin{table}[!ht]
\caption{Cumulative information levels (IL)}
\vspace{-10pt}
\label{tab:levels}
\begin{minipage}{\linewidth}
\begin{center}
\begin{tabu}{lX}
\toprule
\textbf{IL} & \textbf{Description} \\
IL1 & Basic, Length, and Living type categories from Table 1 \\
IL2a & IL1 + binary Time category from Table 1\\
IL2b & IL1 + binary Health Care and Type categories from Table 1\\
IL3 & IL2a+IL2b \\
IL4 & Contains IL3 but with the full distribution (instead of binary) Time, Health Care, Feedback and Financial categories from Table 1\\
\bottomrule
\end{tabu}
\end{center}
\end{minipage}
\end{table}
\subsection{Training and baselines}
%** REWRITE THE FOLLOWING SO THAT YOU CLEARLY DESCRIBE WHAT IS TRAINED AND WHAT ARE THE BASELINES**

The aim is to predict large increases in the hours of homecare provided to citizens. We define large increases as monthly increases of at least 6 hours within the next 3 months. %In order to use the data for predicting future increases in home care time, we choose to 
We aggregate the data over 3-month periods: for each month, we aggregate the home care services of the current month with that of the previous two months. We do this  for each citizen, with increments of 1 month, such that each month occurs in three 3-months periods. %This is done for each month by aggregating the home care services of the current month with the previous 2 months. 
This produces 423,909 historical past records for individual citizens, which can be grouped by month to create the chunks described in section \ref{methodd}. We associate a binary target variable to each record indicating if the provided home care increases by at least 6 hours in the next 3 months, or not. This choice corresponds to 88.05\% negative classes and 11.95\% positive classes.

%We have data available from April 2013 to April 2017, which means that when using 3 months for summarizing a citizen and looking 3 months ahead to determine if the home care service hours were increased by at least 6, then we have training samples available 
Our training data consists of historical past records from September 2013 to February 2017, with %. Each of these training samples contain the 3 month summarization and 
the associated target variables representing if a large increase happens in the following 3 months.
%As described in section 2, w
%We consider monthly chunks of these training samples, and train a level 0 model on each month of data, and level 1 models are trained as explained in section 2. 
We use logistic regression and random forests, presented in section 2, interchangeably as level 0 and level 1 models. For logistic regression, $\lambda$ is tuned in $\lambda \in [10^{-4},..., 10^4]$, testing 100 evenly distributed values in logarithmic scale. For random forests, the number of trees is tuned $\in [100,200,...,1000]$, the number of features in each split is tuned $\in [0.1,0.15,...,0.9]$, and the minimum number of samples for splitting and creating leafs is tuned $\in [2^0,2^1,...,2^{6}]$. At all times, 3-fold stratified cross validation is used for training. At each time point data from 3 months ago is used for training and validation, and tested on the current month.

For level 0 models, we experiment with two variations: 1) using the same level 0 model on all previous available data; and  2) using the data from the last month for the level 0 model. 

%When evaluating the performance of our approach in a given month we only use data up to 3 months before. The reason for this is that since we are looking 3 months ahead, then we can not use models trained on samples from the previous 3 months, as they would contain future information. This means that if we predict on the citizens in August 2016, then we use data available from September 2013 to May 2016, and the 
Performance is evaluated by measuring the area under the receiver operator curve (AUC). AUC was chosen as opposed to e.g. accuracy, because our dataset is highly imbalanced (88.05\% negative classes and 11.95\% positive classes). As AUC measures the trade-off between the true positive and false positive rate, it is not affected by an imbalanced datasest, whereas accuracy is.

%As described in section 2 our proposed method is able to work in a non-stationary environment, and 
By learning the combination strategy of level 0 models we assume that the best prediction is not just based on the most recent model, but found by utilizing multiple previously learned models.

%We take this approach since home care is a domain where changing local and global policies may influence how care takers assign home care, and responsible care takers for each citizen changes over time.

There are no established domain-specific baselines to compare our work to. Even the city of Copenhagen, who provided the data, did not have an automatic or manual method for predicting increases in home care, despite acknowledging the importance of this type of prediction. So we created the following two baselines to evaluate our method: predicting that today's increase in home care hours is the same as 1) that of one year ago; and 2) that of three months ago.

We further experiment with reduced amounts and types of training data, and study their impact on prediction effectiveness. We structure our data into five \textit{cumulative} information levels, shown in Table \ref{tab:levels}.
%, where the highest level is always used if nothing else it mentioned.
%The lowest level consists of standard information such as age, civil status, and where the citizen is living. The highest level contains information on when and what service was provided (on a daily basis), a reason for potential absence (e.g. admitted to the hospital), and who the provider of the service was (public or private). 
%By considering the dataset on different information levels we aim for this work to be applicable even in cities with a limited amount of logging.

\subsection{Findings}
Table \ref{tab:models} displays the AUC of our ensemble method with different combinations of logistic regression (LR) and random forest (RF) as level 0 and level 1 models, as well as the baselines. The AUC values are the average over all months.
%** We said above that one baseline is now=1 year ago, and the other is now=3 months ago. Is that true? If so, where is this in the Table? **
We see that the highest AUC (0.715) is obtained from using logistic regression on all previous training data. This means that isolating only the most recent context is not necessary for this type of prediction. We also see that using logistic regression as both level 0 and 1 learners performs similar with an AUC of 0.714. Thus surprisingly the data stream remained similar during the 5-year period, such that an online ensemble did not provide an improvement.

\begin{figure*}
\includegraphics[scale=0.45]{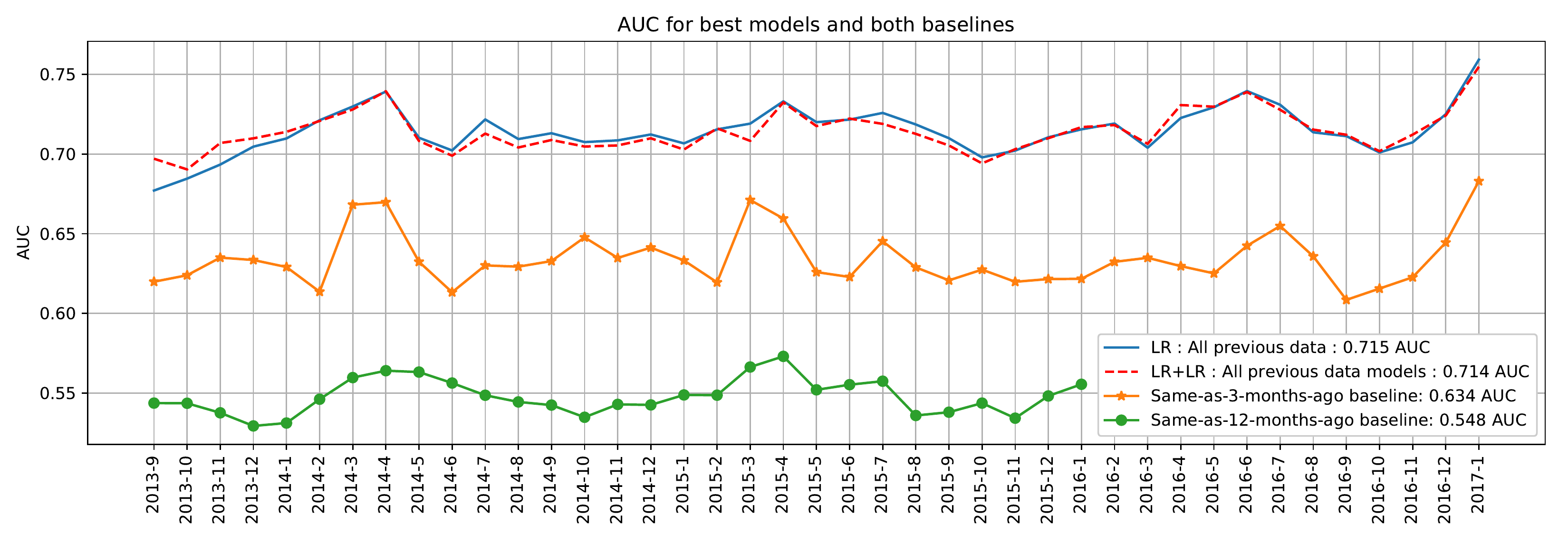}
\vspace{-15pt}
\caption{\label{fig:models} AUC scores for the best two performing models and training data, as well as the baselines. The dashed line correspond to our method.}
\end{figure*}

Figure \ref{fig:models} shows the two best performing models for each month, as well as the two baselines. The best performing methods clearly outperform both baselines, where predicting the same as the year before achieved 0.548, and doing the predictions based on if increases occurred during the last 3 months achieved 0.634.
%We refer to the level 1 model with level 0 models trained on all previous training data as "all previous data models", and the rest follows the same terminology used throughout the text.

\begin{figure*}
\vspace{-15pt}
\includegraphics[scale=0.45]{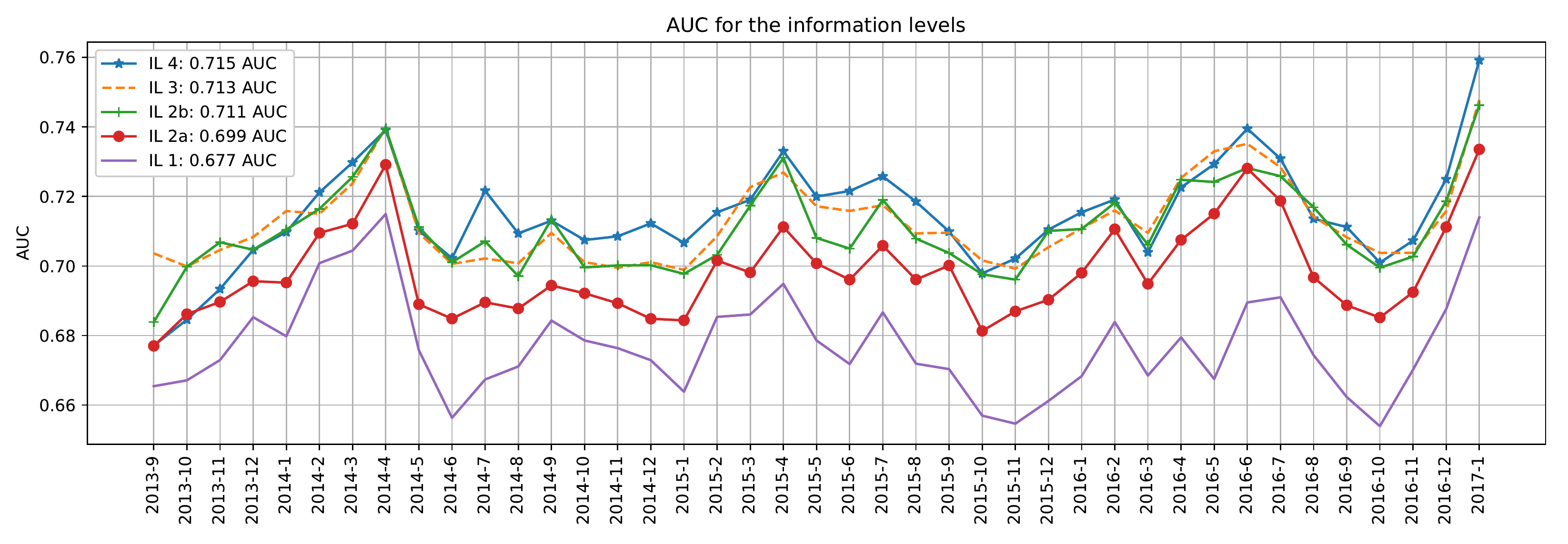} %width=\linewidth
\vspace{-15pt}
\caption{\label{fig:lvls} AUC scores for the best performing models and training data in the five cumulative information levels.}
\end{figure*}

The results when doing the same experiment as above on the five cumulative information levels can be seen in Table \ref{tab:lvls} as averages over all months, and visualized in Figure \ref{fig:lvls} where the best performing models are shown for each month. The best AUC score for the lowest information level was 0.677, which increases to 0.697 at the next level, but the remaining three levels perform very close to each other - however with small increases as the levels increase. All information levels outperform the baselines.
We see that the more information we use for the prediction (increasing Information Levels (IL)), the better the prediction. However, we also see that prediction is not notably lower when using only basic information (IL1). This implies that good predictions can be achieved by using only basic citizen records, which most cities are likely to log.
\setlength{\tabcolsep}{.1666em}
\begin{table}
\caption{Average AUC for models and training data with information level 4 (IL4). The last four columns correspond to our method. Level 1 models using level 0 models trained on last month (1) or all previous data (2) are shown in the last 3 rows.}
\vspace{-8pt}
\label{tab:models}
\scalebox{0.85}{
\begin{minipage}{\linewidth}
%\begin{center}
\begin{tabular}{lccccccc}
\toprule
Trained on/Method & Baseline & RF & LR & RF+LR & RF+RF & LR+RF & LR+LR \\
3 months baseline & .634 & -  & -  &  -    &    -  &    -    &   -     \\
12 months baseline & .548 & -  & -  &  -    &    -  &    -    &   -     \\
(1): Last month & - & .562 & .709 & - & - & - & -  \\
(2): All previous data & -  & .572 & \textbf{.715} & - & - & - & -  \\
Models from (1) & - & - & - & .605 & .578 & .570 & .711  \\
Models from (2)& - & - & - & .617 & .582 & .571 & .714  \\
Models from (1)+(2)& - & - & - & .618 & .575 & .572 & .714  \\
\bottomrule
\end{tabular}
%\end{center}
\end{minipage}}
\end{table}
\setlength{\tabcolsep}{.6em}
\begin{table}
\caption{Average AUC for the best performing model on different information levels (IL) of training data. Level 1 models using level 0 models trained on last month (1) or all previous data (2) are shown in the last 3 rows.}
\vspace{-8pt}
\label{tab:lvls}
\scalebox{0.85}{
\begin{minipage}{\linewidth}
\begin{center}
\begin{tabu}{lccccc}
\toprule
Trained on/Level & IL1 & IL2a & IL2b & IL3 & IL4 \\
(1): Last month     & .669 & .696 & .710 & .711 & .709   \\
(2): All previous data       & .665 & .697 & .709 & .712 & \textbf{.715}   \\
Models from (1)        & .677 & .696 & .711 & .712 & .711   \\
Models from (2)       & .676 & .699 & .711 & .712 & .714   \\
Models from (1)+(2)      & .677 & .698 & .711 & .713 & .714   \\
\bottomrule
\end{tabu}
\end{center}
\end{minipage}}
\end{table}

\section{Discussion and Conclusion}
Our choice of using records from the last 3 months to predict increases in home care was motivated from the assumption that recent policy changes on behalf of the city could have an effect on the amount and level of provided home care. Similarly, changes in who the responsible care taker is could have an effect as well. However, we see that using a model trained on all available data gave the best AUC of 0.715. %This indicates that there is no significant drift in the data, and training on all available data is therefore not problematic.

Our methods were compared against two baselines, one where we predict that an increase will happen if it happened a year before, and another where we predict an increase if it happened within the last 3 months. These baselines serve as heuristic real-world ways of planning ahead. With baseline AUC scores of 0.548 and 0.634 respectively, we were able to outperform them by a large margin, thus providing a considerable improvement.

When inspecting the weights of the model trained on all previous data in the last time step we find that the five most discriminative features and their absolute weights are: the number of large increases in the last 3 months (0.38), if the citizen was hospitalized (0.22), if the citizen received sick care (0.21), if the citizen received home care in the weekend (0.21), and the age of the citizen (0.14). Interestingly the most important feature, by a large margin, was the number of times a citizen had 
their home care hours largely increased in the past 3 months, which corresponds to the best performing baseline. This could correspond to an initial under-evaluation of the new level of required home care for a citizen when their circumstances change, or that a decline in a citizens' capabilities happens fast over a relatively short time span.
 
We argued in section 1 that the problem of predicting large increases in home care is similar to that of predicting hospital re-admissions, where recent model comparisons shows the highest AUC scores of 0.68 to 0.73 \cite{tong2016comparison, futoma2015comparison} depending on the data. The purpose of those studies has been to obtain financial savings and to put extra focus on patients likely to be re-admitted. The purpose of our work is similar, and may be used as a tool for long term planning for both financial and health related benefits. 

We considered our data at five cumulative information levels in order to understand what level of detail is needed to obtain good performance. While we had very detailed data available as described in Table \ref{tab:data}, we found that simply using binary information about the received services on each citizen (IL2b) performed similar to using all available information, which is beneficial since this data is extremely cheap to produce since no daily logging is needed. The binary information about the received services correlates with the general health of citizens, and can therefore also be seen as a derived feature based on their actual medical conditions. Since the largest performance increase comes from knowing which services the citizens' receive, a very plausible hypothesis is that knowing the actual conditions of each citizen would lead to further performance improvement. However, this kind of information is not always easily accessible.

This has been a preliminary study into smart city analytics, where we focused on using a well known machine learning methodology for making predictions, namely ensemble learning, in order to predict one of the major health expenses for a city: large increases in home care hours received by citizens. In the future we plan to  use  natural language processing in order to extract meaningful semantics from the journals of the carers (e.g. nurses, doctors, or other staff) and investigate its usefulness to smart city predictions that can improve welfare.
This work complements
wider efforts in smart health analytics using Machine Learning
\cite{HansenMCL17,HansenMCL17s}, and Ensemble Learning \cite{HansenLM16} in
particular.

%As future work there are various paths to explore. First, in this work we aggregated historical data to create a feature vector, however, it could be interesting in the future to use the data directly through a recurrent neural network approach or using the existing representation through other ensemble based techniques \cite{gama2014survey}. Second, incorporating more data sources, where we are currently working on getting access to a semi-structured textual log of each citizens' well being.
\begin{acks}
Funded by the Innovation Fund Denmark, DABAI project.
\end{acks}
\bibliographystyle{ACM-Reference-Format}
\bibliography{../sigproc} 
\end{document}